\documentclass[10pt,romanappendices,conference,twocolumn]{IEEEtran}
\usepackage{cite}

\usepackage{amssymb}
\usepackage{graphicx}
\usepackage{amssymb}
\usepackage{amsbsy}
\usepackage{float}
\usepackage{balance}
\usepackage{subfigure}
\usepackage{url}
\usepackage{bm}
\newfloat{longequation}{b}{ext}
\newfloat{longequation}{t}{ext}

\usepackage[cmex10]{amsmath}

\usepackage{array}
\usepackage{varioref}
\usepackage[normalem]{ulem}

\graphicspath{{./figs/}}
\usepackage{algorithm2e}
\DeclareMathOperator{\Tr}{Tr}
\DeclareMathOperator{\diag}{diag}

\begin{document}

\title{Constructive Interference   in  Linear Precoding Systems: Power Allocation and User Selection}

\author{\IEEEauthorblockN{ Dimitrios Christopoulos\IEEEauthorrefmark{1}, Symeon Chatzinotas\IEEEauthorrefmark{1}, Ioannis Krikidis\IEEEauthorrefmark{2}   and Bj\"{o}rn Ottersten\IEEEauthorrefmark{1}}
\IEEEauthorblockA{\IEEEauthorrefmark{1}SnT - securityandtrust.lu,  University of Luxembourg
\\e-mail: \textbraceleft dimitrios.christopoulos, symeon.chatzinotas,  bjorn.ottersten\textbraceright@uni.lu}
\IEEEauthorblockA{\IEEEauthorrefmark{2}Department of Electrical and Computer Engineering, University of Cyprus \\ e-mail: krikidis.ioannis@ucy.ac.cy\   }

}
\maketitle


\begin{abstract}
The exploitation of interference in a constructive manner has recently been proposed for the downlink of multiuser, multi-antenna transmitters.
This novel linear precoding technique, herein referred to as constructive interference zero forcing (CIZF) precoding, has exhibited substantial gains over conventional approaches; the concept is to cancel, on a symbol-by-symbol basis, only the interfering users that do not add to the intended signal power. In this paper, the power allocation problem towards maximizing the performance of a CIZF system with respect to some metric (throughput or fairness) is investigated. What is more, it is shown that the performance of the novel precoding scheme can be further boosted by choosing some of the constructive  multiuser interference terms in the precoder design.   Finally, motivated by the significant effect of user selection on conventional, zero forcing (ZF) precoding, the problem of  user selection for the novel precoding method is tackled. A new iterative, low complexity algorithm for user selection in CIZF is developed. Simulation results are provided to display the gains of the algorithm compared to known user selection approaches.
\end{abstract}
\section{Introduction }
The capacity of the multiple input multiple output (MIMO) broadcast channel (BC) can be reached by non-linear precoding methods, namely dirty paper coding (DPC)\cite{Costa1983}. However, linear precoding methods, like zero forcing (ZF) precoding, can still attain the channel capacity in a multiuser environment \cite{Yoo2005,Yoo2006b,Ng_TIT_08}, while proven more realistic in terms of practical  implementation. Linear precoding techniques, especially ZF, have been extensively investigated  in \cite{Wiesel-08,Yoo2006b} and the references therein. In these cases, ZF precoding constitutes a simple precoder design solution. By inverting the channel,   multiuser interferences are cancelled and the precoding design problem is reduced to a  power allocation problem over new equivalent channels; hence a simple concave optimization problem\cite{dimic2005} needs to be solved. To maximize the  throughput (sum-rate, SR), the well known water-filling solution can be straightforwardly applied \cite{Yoo2006}. To maximize the minimum offered rate (i.e. the fairness problem), the problem is still convex and thus solvable\cite{Wiesel-08}.
 The key assumption of all the above considerations however is the assumption of Gaussian signaling.

\begin{figure}[ht]
\centering
\includegraphics[width=1\linewidth]{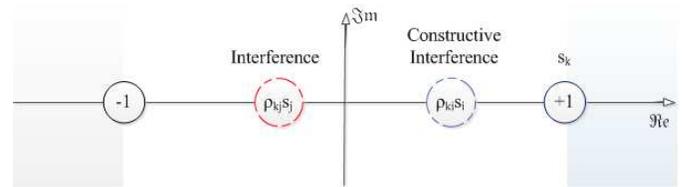}\\
\caption{The constructive interference (CI) concept over binary phase shift keying (BPSK) modulations: The $k$-th user's transmit symbol is $s_k=+1$. The $i$-th user's symbol, $s_i$, multiplied by the cross-correlation between the $k$-th and the $j$-th user's channels, $\rho_{kj}$ (see Sec. \ref{sec: CIZF}), is a vector that when added to $s_k$ will move the resulting vector further away from the decision threshold (0 for BPSK). Subsequently, not cancelling this user will benefit the  $k$-th user. On the other hand, the $j$-th user is still interfering thus needs to be cancelled by the precoding design.  }\label{fig CI concept}
\end{figure}

The concept of constructive interference  linear precoding, initially proposed  in \cite{Masouros2007} for code division multiple access (CDMA) systems and then extended to apply for MIMO communications in \cite{Masouros2009}, is based on the multiuser interference cancellation concept of channel inversion. An example of the concept is described in Fig. \ref{fig CI concept}.
The novelty of this precoder design  lies in considering  practical constellations and  allowing users that add up to the intended user's signal power to interfere. This is referred to as constructive interference (CI) and it can be exploited by acknowledging each users' channel  and modulated signal.
The problem of power allocation in constructive interference zero forcing (CIZF) precoding techniques has not been studied in existing literature. Existing works on this topic only assumed CIZF precoding with equal power allocation for all users\cite{Masouros2009}.

 Another very important aspect of linear precoding is the user selection problem investigated in \cite{Yoo2006b,dimic2005}.  ZF  performance is increased when user channels are  orthogonal to each other. Under the assumption of large random user sets, the probability of orthogonal users increases and with that the complexity of the user selection problem. Nevertheless, simple suboptimal  algorithms in the existing literature provide substantial gains with affordable complexity. Based on existing methods, \textit{Yoo et al} \cite{Yoo2005} proposed a low complexity, iterative user selection algorithm that allows ZF to achieve the performance of non-linear precoding\cite{Weingarten2006a}  as the number of available for selection users grows to infinity.

The contribution of the present paper is twofold.  Firstly,   the effect of power allocation on CIZF precoding transmitters   is investigated; a design parameter that has not been examined in existing literature. Secondly, motivated by the fact that user selection can optimize the ZF performance, the problem of user selection in CIZF systems is defined and solved by a low complexity algorithm. This algorithm achieves substantial gains and approaches the performance of the optimal user selection, as derived by  full space search.

The rest of the present paper is structured as follows. The considered system model is described with detail in Section \ref{sec: System model}, where 
the concepts of conventional ZF and novel CIZF are also described. Section \ref{sec: Power control} explores the effects of power allocation in CI based linear precoding methods with the support of simulation results.  In Section \ref{sec: User selection}, the user selection problem is defined and solved via a novel heuristic algorithm. Conclusions are drawn in Section \ref{sec: conclusions}.



\textit{Notation}: Throughout the paper,  \(\left(\cdot\right)^\dag\), $\mathfrak{Re}(\cdot)$ and \(||\cdot||, \) denote   the conjugate transpose, the real part of  complex elements and the Euclidean norm operations, respectively, while $[\cdot]_{ij}  $  denotes the $i, j$-th element of a matrix. The element-wise matrix product is denoted by $\circ$.
Bold face lower case characters denote column vectors and upper case denote matrices while the operator diag($\mathbf x$) produces a diagonal square matrix composed of  the elements of  $\mathbf x$. An identity matrix of size  $n $ is denoted by $\mathbf{I}_n$.  Upper case calligraphic characters denote sets. The operation $\mathcal {A-B}$ is the relative complement of $\mathcal B$ in $\mathcal A$, while   $|\mathcal A|$, denotes the cardinality of a set.
\section{System Model}\label{sec: System model}
A multi-user (MU) multiple input single output (MISO) network consisting of one transmitter with $N_{t}$ antennas and $K\geq N_{t}$ single antenna receivers, is considered. At each time, the transmitter serves exactly $K = N_{t}$ users which are selected either randomly or based on a selection scheme as described in Sec. \ref{sec: User selection}.  The received signal at the $k$-th user can be expressed as
\begin{equation}\label{eq: General input-output}
y_{k}= \mathbf h^{\dag}_{k}\mathbf x+n_{k},
\end{equation}
where \(\mathbf h_{k}\) is an \(N_{t} \times 1\) vector composed of the channel coefficients between the \(k\)-th user and the  \(N_{t}\) antennas  of the source, \(\mathbf x\) is an \(N_{t} \times 1\)  vector of transmitted symbols and  \(n_{k}\) is the independent identically distributed (i.i.d) zero mean  Additive White Gaussian Noise (AWGN)  measured at the \(k\)-th user's receive antenna. The noise is assumed normalized, thus $ \mathcal E\left\{|n_k|^2\right\}=1 $. In matrix form,
this MU MISO BC reads  as
\begin{align}
\mathbf{y}=\mathbf{H}\mathbf{x}+\mathbf{n},
\end{align}
where $\mathbf{y}=[y_1,y_2,\ldots,y_{N_t}]^\dag$, $\mathbf{x}=[x_1,x_2,\ldots,x_{N_t}]^\dag$, $\mathbf{H}$ is the $N_t\times N_t$ square matrix that contains the   user complex vector  channels, i.e. \(\mathbf H = \left[\mathbf h_1 ,\mathbf h_2\dots,  \mathbf h_{N_t} \right]^\dag\) and $\mathbf{n}=[n_1,n_2,\ldots,n_{N_t}]^{\dag}$. The transmitter  linearly precodes the information symbols:
\begin{align}\label{eq: precoding}
\mathbf{x}=\mathbf{W}\mathbf{P}^{1/2} \mathbf{s},
\end{align}
where the $N_t\times N_t$ matrix $\mathbf{W}$ is the precoding matrix, $s_{k}$ denotes the symbol for the $k$-th destination and $\mathbf{s}=[s_1,\ s_2,\ldots,s_K]^\dag$ with $\mathbb{E}[\mathbf{s}\mathbf{s}^\dag]=\mathbf I$ and $\mathbf{P}^{1/2}=\diag([\sqrt {p_{1}}, \sqrt {p_{2}} \ldots \sqrt {p_{k}}])$ is a diagonal   $K\times K$ matrix composed of the transmit powers allocated to the $k$ users\footnote{The notion of transmit power allocated to a user is explained in \cite{Yoo2006b}.}.
For shortness and since the results can be straightforwardly generalized for higher order constellations, in this study only real valued signals will be assumed (binary phase shift keying-BPSK modulation), hence
\begin{align}
{s_i}=\pm 1,\  i=1\ldots N_{t}.
\end{align}

\subsection{Zero Forcing beamforming}\label{sec: conv ZF}
Transmit beamforming is a multiuser precoding technique that separates user data streams in different parallel beamforming directions\cite{Yoo2006b}. A linear precoding technique with reasonable computational complexity that still achieves full spatial multiplexing and multiuser diversity gains, is ZF precoding \cite{Viswanathan2003,Caire2003,Yoo2006b}. The ability of ZF to fully cancel out multiuser interference makes it useful in the high Signal to Noise Ratio ($\mathrm {SNR}$) regime. However, it performs far from the optimal in the noise limited regime. In addition,  it can only simultaneously  serve  as many  single antenna users as the number of transmit antennas. A common solution for the ZF precoding matrix is the pseudo-inverse of the $K\times N_{t}$ channel matrix.
 Under a total power constraint, the pseudo-inverse  is the optimal solution (rather than any generalized inverse) in terms of maximum SR and maximum fairness \cite{Wiesel-08}.
The precoding matrix can be expressed  as
\begin{align}
&\mathbf{W}= \mathbf{H}^\dag \mathbf{R}^{-1}\mathbf{T},\label{eq: Gen Precoder}
\end{align}
where we define the matrix $\mathbf R$  as
\begin{align}\label{eq: R}
 \mathbf R = \mathbf{HH}^\dag.
 \end{align}
%
 The matrix $\mathbf T$ has been introduced by \cite{Masouros2009} to model the CI scheme as explained in the next subsection. 
The general model described by \eqref{eq: Gen Precoder}, for
\begin{align}
\mathbf{T}=\mathbf{I} \circ \mathbf{R,}
\end{align}
where the non zero elements of $\mathbf T$ are $\tau_{kk}=\sum\mathbf h_k^\dag \mathbf h_k$, will yield the conventional ZF design.  The complete cancellation of interferences in this case, will reduce the $k$-th users received signal to
\begin{align}\label{eq: General input-output2}
&y_{k}= \mathbf h^{\dag}_{k}\sqrt{p_k}\mathbf w_{k} s_{k}+ n_k, 
\end{align}
where $\mathbf w_k$ is the $k$-th column of the total precoding matrix. Assuming uniform power allocation across users, as in \cite{Masouros2009}, the total power constraint over the transmit antennas $P_{tot}$ will yield\cite{Wiesel-08}:
\begin{align}
\mathcal E\{||\mathbf x||^2\}= \Tr\{\mathbf {x x}^\dag \}\leq P_{tot}.
\label{eq: Sum Power Constraint}
\end{align}
From \eqref{eq: precoding} and \eqref{eq: Sum Power Constraint}, the transmit power allocated to the $k$-th user becomes

\begin{align}
\sqrt p_k= \sqrt{\frac{P_{tot}}{\Tr\{\mathbf {T}^\dag \mathbf R^{-1} \mathbf{T} \}}} , \  \forall \  k.\label{eq: Pk}
\end{align}
By  examining \eqref{eq: Pk}, it is clear that the transmit power allocated
to the $k$-th user is a function of the precoder design. In precoding, channel inversion i.e. projecting the actual channels on orthogonal dimensions, leads to the reduction of each users effective channel. Subsequently, since the precoders are not normalized,  the sum of the individual powers   allocated to each user ($\sqrt{p_k}$) is not equal to the sum of powers transmitted \eqref{eq: Sum Power Constraint}. The notion of individual user consumption can be introduced to better explain this power loss due to the precoding. Finally, the    $k$-th user $\mathrm{SINR}$ for the ZF precoding will read as
 \begin{equation}\label{eq: SINR ZF}
\mathrm {SINR}_k^{\text{ZF}}= |\tau_{kk}|^2 p_k.
\end{equation}

\subsection{Constructive Interference Zero Forcing beamforming}\label{sec: CIZF}
The CIZF scheme, introduced in \cite{Masouros2009}, allows the so-called constructive  multiuser interference (cross-interference) to be added to the useful signal at each receiver.
In general, given the full channel state information available at the transmitter and acknowledging the signal constellation,  the CIZF scheme does not suppress the part of the cross-interference that is constructive and thus increases the power of the useful signal. A simple example of this concept is explained in Fig. \ref{fig CI concept}.

As discussed with detail in \cite{Masouros2009}, the symbol to symbol multiuser interference results from the $i,j$-th element of the  matrix $\mathbf R$ : $\rho _{ij}=\sum_{n=1}^{N_t} h_{in} \cdot ({h_{jn}}^\dag)$. In the CI scenario, the received signal of \eqref{eq: General input-output2} will become
  \begin{align}\label{eq: CI y}
 y_k=\mathbf  \tau_{kk} \sqrt{p_k}s_k+\sum_{j\neq k}\text{CI}_{kj} +n_{k}
 \end{align}
 where  $\text{CI}_{kj}=\tau_{ki}  \sqrt{p_j}s_j$ denotes the constructive cross-interference from the $j$-th data flow ($j$-th user) to the $k$-th user.
Subsequently, the $k$-th user's signal to interference plus noise ratio (\( \mathrm{SINR}\)) will read as
 \begin{equation}\label{eq: SINR}
\mathrm {SINR}_k=\sum^{K}_{j=1
} |\tau_{kj}|^2 p_j.
\end{equation}
 Let us define as
 $\mathbf G = \text{diag}(\mathbf s)\cdot \mathfrak{Re}(\mathbf R)\cdot \text{diag}(\mathbf s), $ which yields
\begin{align}\label{eq: CI matrix}
\mathbf{G}=
\begin{pmatrix}
s_1 \mathfrak{Re}( \rho_{11}) s_1 & \dots &  &  \\
 \vdots & \ddots & & \\
 &  & s_{k} \mathfrak{Re}( \rho_{kj}) s_{j} &  \\
 &  &  & \\
\end{pmatrix}.
\end{align}
In order to indicate the cross-interference as CI, the signal constellation needs to be accounted for. Subsequently, the terms that position the received signal into the decision region of the transmitted symbols are beneficial and thus not cancelled by the precoding design. For the simple case of BPSK modulation, the cross-interference generated by the $j$-th data flow to the $k$-th destination, is considered to be constructive when
\begin{align}
s_k \mathfrak{Re}(\rho_{kj}) s_j >0,
\end{align}
which  can be expressed as $\mathbf G_{kj} >0.$
Thus the CIZF precoder is deduced from
\begin{align}
&\tau_{kk}=\rho_{kk} \\
&\tau_{ki}=
\begin{cases}  \rho_{ki}, & \text{If}\ [\mathbf G]_{kj} > 0   \\
0,&\text{elsewhere.}\notag    
\end{cases}
\end{align}
Therefore, the precoding matrix is computed on a symbol-by-symbol basis.

\section{Power Allocation in Linear Precoding}\label{sec: Power control}

The impact of power allocation  (PA) on the CIZF has not been addressed in existing literature on this topic \cite{Masouros2007,Masouros2009}. Therein, the problem was simplified by a uniform power allocation assumption,  as defined in \eqref{eq: Pk}. In general, PA is performed to the end of maximizing some performance metric. The performance metrics commonly addressed in  literature involve either the total throughput performance (i.e. max SR criterion)  or the \( \mathrm{SINR}\) level of the worst   user (i.e. max fairness criterion).  Another important parameter in linear precoding is the type of constraints that will be assumed. Usually, a total sum power constraint simplifies the analysis and provides better results since the available power is freely allocated across antennas. Herein,  two objective functions of the achievable user rates that ensure maximum fairness (availability) and maximum SR (throughput), are considered. More specifically, the optimization problem reads as
\begin{align}\label{eq: max SR PA}
&\max_{\mathbf{P}\geq \mathbf{0}} f(\mathbf{P}) \\
\text{s.t.}\ & \sum_{i=1}^{N_t}\sum_{j=1}^{K}|\omega_{ij}|^2 p_j \leq P_{tot} \nonumber
\end{align}
where $\omega_{ij}$ is the $i, j$-th element of $\mathbf W$ and  the objective function $f$ is given by \cite{Wiesel-08}:
\begin{align}\label{eq: opt}
f(\mathbf{P})=
\begin{cases}
\sum_{k} \log_2(1+\mathrm {SINR}_k),& \text{\ Throughput}\\
\min_{k} \mathrm {SINR}_k, &\text{\ Fairness}
\end{cases}
\end{align}
where $\mathrm {SINR}_k$ is given by \eqref{eq: SINR} and $1\leq k\leq N_{t}$.
Based on \eqref{eq: opt}, the objective function is concave in $\mathbf{P}\geq 0$, for both scenarios and therefore the optimization problem is a simple concave maximization with one linear constraint.  It is worth noting that for the case of the maximum throughput, the optimization problem can be solved using the water-filling solution.
 The problem of allocating the power to the end of maximising some system performance metric is discussed in the following Section (\ref{sec: PA prob}).

\subsection{Power Allocation }\label{sec: PA prob}
An appropriate PA scheme distributes the total available power to the data flows in a way that maximizes an objective function of the achievable rates. In the case of conventional ZF precoding, the max throughput PA problem, under a total power constraint, reads as in \eqref{eq: max SR PA} with objective function
\begin{align}\label{eq: ZF PA}
f(\mathbf P)=\sum_{k} \log_2\left(1+\mathrm {SINR}_k^{\text{ZF}}\right),
\end{align}
 where the $\mathrm {SINR}_k^{\text{ZF}}$ is given by \eqref{eq: SINR ZF}.

In order to be able to easily show the impact of PA on CIZF and maintain concavity for the formulated optimization problems, we assume that the equivalent CIZF channel (with the modulation-based CI) refers to Gaussian inputs; this assumption allows to approximate the channel capacity of the system with the simple $\log$-based Shannon expressions. It should be clarified here, that the optimal power allocation problem for linear precoders under the constraint of finite input alphabets is a highly complex problem. The most recent attempt to solve it can be found in \cite{Wu2012} where a  heuristic optimization algorithm is developed. However, in the present paper, a preliminary study to exhibit the impact of PA on CIZF is performed, hence the strictly optimal solution is beyond the scope of the present work.

\subsubsection{Simulation Results} 
\begin{figure}[ht]
\centering
\includegraphics[width=1\linewidth]{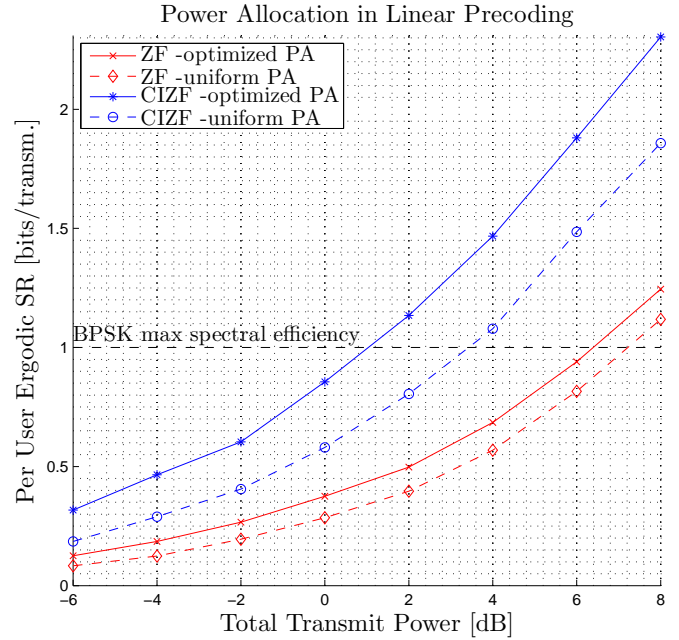}\\
\caption{Per user spectral efficiency of conventional ZF precoding under uniform and optimized PA, compared to the efficiency of the novel CIZF precoding under equivalent PA assumptions. PA optimization for max system throughput. The modulation constrained threshold for BPSK is also plotted.}
\label{fig: PA res 1}
\end{figure}

The effect of power allocation in the CIZF precoding design is plotted in Fig.  \ref{fig: PA res 1}. The power allocation problem \eqref{eq: opt} has been solved using the CVX tool in MATLAB \cite{convex_book}. Simulations where carried for $100$ channel instances, and for $K =N_{t} = 4$.    In Fig. \ref{fig: PA res 1}, the  gain from CIZF   with uniform power allocation, as proposed in \cite{Masouros2009}, over the conventional ZF  is evident by comparing the dashed lines.      The novel result, depicted in Fig. \ref{fig: PA res 1}, is that power allocation further boosts the CIZF gain (continuous lines). More specifically, for the conventional ZF scheme power allocation introduces approximately $1$ dB of gain over the uniform  allocation. However, when PA is applied in CIZF, then more than $2$ dB  gain can be gleaned. It is therefore concluded that power allocation over the CIZF precoding scheme is an important aspect that introduces significant gains.
Finally, in the same figure,  the realistic region where the  results apply is defined by a dashed line. This restriction comes from the acknowledgement of BPSK modulation as a practical constellation choice.
The alleviation of this restriction via adaptive modulation methods is part of the future extensions of this work.

\subsection{Power Constrained Transmission }\label{sec: optimal CIZF}

In the present section, the existence of redundant CI interference terms is discussed.  As can be seen from \eqref{eq: Pk} and \eqref{eq: SINR}, the consideration of non-zero, off diagonal elements in the precoding matrices, i.e.\ CI  terms, has a double impact on each user's $\mathrm{SINR}$ and thus the sum system capacity; from one hand, it increases the expression $\sum_{i} |\tau_{ki}|^2   p_k$ by adding more positive terms in the summation, but on the other hand, it changes the power allocated to each user in \eqref{eq: Sum Power Constraint}, \eqref{eq: Pk}. Therefore,  a CI term in the CIZF is not always beneficial for the system performance. This partially constructive interference zero forcing (P-CIZF)  scheme examines the tradeoff between the positive and the negative impact of an CI term by searching all combinations and selecting the most beneficial set of CI terms. The P-CIZF scheme can be formulated as
\begin{align}\label{eq: PCIZF}
&\max_{\{\mathbf p, \mathbf T^{(\mathcal S)}\}} f(\mathbf{p},\mathbf T^{(s)})  \\
\mbox{s.t.}\ \ & \mathbf T^{(\mathcal S)} \subseteq \mathbf T^{(\mathcal S_{tot})},\notag\\
  & \sum_{i=1}^M  \sum_{j=1}^K| \omega_{ij}|^2 p_{j}   \leq P_{tot}. \label{eq: opt PA}
\end{align}
\noindent where
\begin{align}
\mathcal S \subseteq \mathcal S_{tot} =2^{m}, \nonumber
\end{align}
with $m =  \left| [\mathbf G]_{ki} > 0 \right|$, the number of CI terms. The above definition means that the Partially-CIZF scheme searches all the possible combinations ($2^m$) of the CI terms and holds the one that maximizes the objective function considered.
The optimization problem \eqref{eq: PCIZF}, will be solved under uniform \eqref{eq: Pk}  and optimal \eqref{eq: opt PA}  power allocation considerations. Intuitively, the second consideration provides more flexibility in the design but finding the strictly beneficial terms while at the same time optimizing the PA is a highly complex procedure;  for every possible combination of the CI terms, a new convex optimization PA problem is solved.

  As the purpose of this work is to demonstrate this interesting trade-off, more practical implementation of the P-CIZF scheme are beyond the scope of this paper.
\subsubsection{Simulation Results}
\begin{figure}
\centering
\includegraphics[width=1\columnwidth]{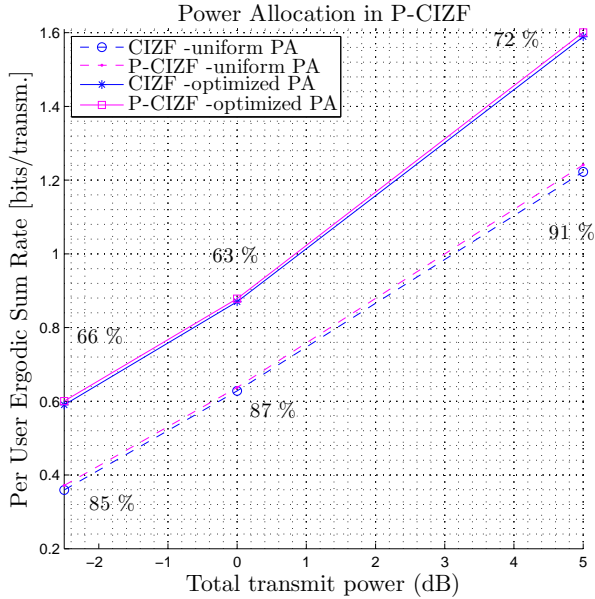}\\
\caption{Evaluation of the P-CIZF compared to fully CIZF under equal and optimal power allocation. The percentages of the constructive terms maintained in the P-CIZF scheme,  for each $\mathrm{SNR}$ point are also presented. }\label{fig: PA res 2}
\end{figure}

In Fig. \ref{fig: PA res 2} the  improvements of P-CIZF over the CIZF scheme are depicted.  The performance is evaluated under uniform and optimized power allocation.  Starting with the CIZF scheme under uniform power allocation, in Fig. \ref{fig: PA res 2}, finding the strictly beneficial CI terms provides some gains. In each point of the figure, the percentage of the CI terms kept in the P-CIZF scheme over the total CI terms of the CIZF scheme is also depicted. Focusing on the dotted curves, in the uniform power allocation case, it is apparent that maintaining approximately $88\%$  of the total number of  CI terms, provides small gains. These   results exhibit the optimality of this approach, while analytical proof of the optimality is part of future work.
The continuous curves in Fig.  \ref{fig: PA res 2} correspond to an optimal (with respect to maximizing the total SR) power allocation assumption in  P-CIZF precoding and again some gains are gleaned; thus the strict optimality of this approach is exhibited. Since the degrees of freedom in the precoder design are increased, less CI terms are maintained (approximately $65\%$).

Intuitively, the above results can be explained by the fact that allowing only strictly beneficial terms in the precoder reduces power consumption, thus allowing for more power to be allocated to the users without exceeding the total power constraint imposed on the transmit antennas.   It should be noted here, however, that the effects of this approach on the minimum supported rate  (fairness) are not examined in the present work. Finally, higher gains of this approach could be gleaned over larger user sets, but by exponentially increasing the iterations of the searching algorithm. The investigation of such scenarios is part of future work.


\section{User selection}\label{sec: User selection}
In this section, the user selection problem is formulated and a novel selection technique based on the exploitation of CI is proposed. Considering the decoupled nature between the user selection and the PA problem, as proven in \cite{Yoo2006b}, the performance of user selection is not affected by PA. However, in accordance with the previous results and more importantly to maximise the overall performance of the assumed system, PA is optimized independently  after the user selection process.
\subsection{User Selection Methods}

\subsubsection{Orthogonal user selection} ZF beamforming has the potential of approaching the optimal channel capacity, otherwise only achievable by DPC, when all the users are perfectly orthogonal to each other, as proven by \cite{Yoo2006b}. The same authors provided a heuristic low complexity user selection algorithm, namely the\textit{ semi-orthogonal user selection} ($\mathrm{SUS}$) algorithm, which was proven to select the optimal users, as the number of available users  approaches infinity.
However, it has never been applied in the CIZF framework. Due to lack of space, the reader is referred to\cite{Yoo2006b,Yoo2005} for more details on this algorithm.  It should be mentioned here, that  the orthogonal user selection does not take into account the CI during the selection procedure. However, to maintain fairness in the study, after the selection, any CI terms that exist are maintained.

\subsubsection{ Optimal Constructive Interference User Selection} The investigated precoding scheme strongly depends on the transmitted signal (user constellation) and therefore the above conventional user selection schemes become unsuitable. A CIZF-based user selection metric should also consider the transmitted symbols, since the precoding matrix is defined via the CI and affects the final performance.
%
In order to attain an upper bound for any CIZF-based user selection policy, all the possible combinations of users $\mathcal Q$ are examined and out of them, the combination $\mathcal U$ that maximizes the system throughput is chosen:
\begin{align}
&\mathcal U=\arg_{\mathcal U}\max_{\mathcal U \in \mathcal Q} \sum_{m}  \log_2 \left(1+\mathrm{SINR}_m\right),
\end{align}
where $\mathrm{SINR}_m$ is given by \eqref{eq: SINR} under a uniform power allocation assumption, i.e. \eqref{eq: Pk}.  It should be stressed that this method relies on exhaustive search over all possible combinations of users.  As a result,  a searching algorithm requires $ \binom{K}{N_{t}} = K!/(K-N_t)!$ iterations in order to decide about the optimal  combination at each transmission. Considering that user selection methods perform better as the number of users increases\cite{Yoo2006b}, i.e. as $K>>N_{t}$, then  the optimal solution becomes difficult to compute. In the scope of this work, a simpler heuristic algorithm is presented hereafter.  \subsubsection{Semi parallel user selection} Inspired by the concept of user orthogonality, the purpose of this selection method is dual: users with CI need to be selected and furthermore these users need to be aligned (rather than orthogonal) so that the aggregate beneficial receive power is increased. Following this concept  the \textit{semi-parallel user selection} ($\mathrm{SPUS}$) algorithm, provided in pseudo-code  in Alg. \ref{fig: SPUS}, has been developed. An analytic description of the algorithm follows.

Initially, the algorithm accepts as input the CI matrix $\mathbf G $. The first step is to choose the user with the larger diagonal element. For this user, the cross-correlation elements with all other users are stored in the buffer vector $\mathbf c_{(i)}$ where $i$ is the iteration counter. Also the sets $\mathcal S, \mathcal T$ that include the available and the selected users, are initialized and updated in every iteration. Cross-correlation is the inner  product of the vector channels of  two users and represents the level of orthogonality between  the users. In this scenario, the purpose is to have as little orthogonality as possible so as to increase the received CI.   In each of the $M$ iterations, (Step 2) the user with the strongest element in  $\mathbf c_{(i)}$ is chosen. Then this user's corresponding cross-correlations with all the users is added in the buffer matrix.
By adding the cross-correlation of each selected user in the buffer matrix, a metric for the  subspace of the previously selected users is created, since the aim of the algorithm is to find the most parallel users to the subspace spanned by the ones already selected. This is achieved by choosing the strongest element of the $\mathbf c$ vector in each iteration.  Since  in the previous steps no guarantee exists that a selected user has only CI towards the selected ones, in Step 3, the residual non-CI terms are removed from the precoding matrix.
 The developed heuristic, iterative algorithm runs for exactly $K$ iterations. 
\subsubsection*{Simulation Results} 
\begin{figure}
  \centering
  \includegraphics[width=1\columnwidth]{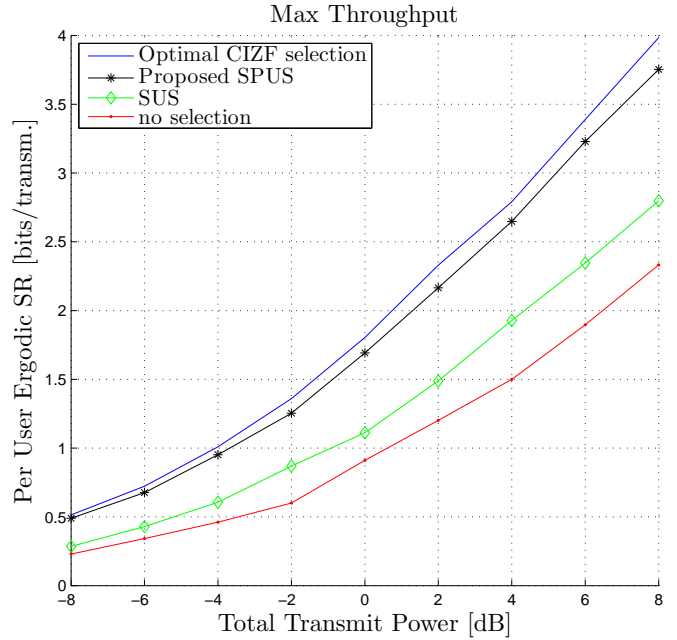}\\
  \caption{Performance of user selection algorithms with respect to the total available transmit power. Results for $N_t = 4$ users selected out of a total pool of  $K = 12$ users. PA has been optimized to maximise the total system throughput.}
\label{fig: GlobeCom 1}
\end{figure}

A comparison in terms of maximum SR performance of the algorithms described in the previous section is presented in Fig. \ref{fig: GlobeCom 1}, where the performance of these algorithms was studied with optimized PA to maximize the total throughput.  In this figure, the gain of the optimal user selection algorithm compared to existing approaches is clear.  The best algorithm for ZF precoding, i.e. $\mathrm{SUS}$ \cite{Yoo2006b}, performs  better than assuming no selection; however, approximately $6$ dB loss is expected over the optimal CIZF selection. This observation emanates the need for better user selection algorithms, when exploiting the benefits of CI.  Accounting also for the complexity of an exhaustive search selection algorithm, as discussed in the previous section, a less complex heuristic algorithm is further necessitated. In this direction, $\mathrm{SPUS}$ has been developed. In Fig.  \ref{fig: GlobeCom 1}, the close to optimal performance of the developed  algorithm is clear; the proposed technique performs less than 1 dB away from the optimal selection.

\begin{figure}
  \centering
  \includegraphics[width=1\columnwidth]{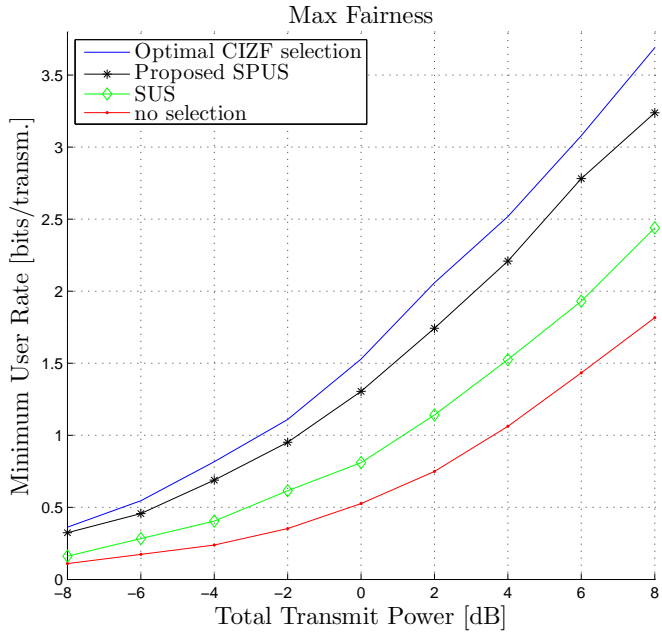}\\
  \caption{Performance of user selection algorithms with respect to the total available transmit power. Results  for  $N_t = 4$ users selected out of a total pool of  $K = 12$ users. PA has been optimized to maximise minimum user rate. }
\label{fig: GlobeCom 3}
\end{figure}

Furthermore, the performance of the developed algorithm under fairness maximization PA is examined in Fig. \ref{fig: GlobeCom 3}. Results indicate, that   when the $\mathrm{SPUS}$ algorithm is combined with max fairness PA optimization, the performance degradation with respect to the optimal selection policy, is relatively small. Therefore, the proposed algorithm, developed for maximising the total throughput of the system, does not severely compromise the fairness of the system.  By comparing Fig. \ref{fig: GlobeCom 1} and \ref{fig: GlobeCom 3}, it is also noted that the   the minimum user rates are not far from the average rate. This result indicates that the variance between the user rates is kept in reasonable levels when user selection is combined with PA to optimize the total  SR. It is therefore concluded that fairness is not severely compromised\ in user selection scenarios, even when PA optimization is performed to maximize to total system SR.

\begin{figure}
  \centering
  \includegraphics[width=1\columnwidth]{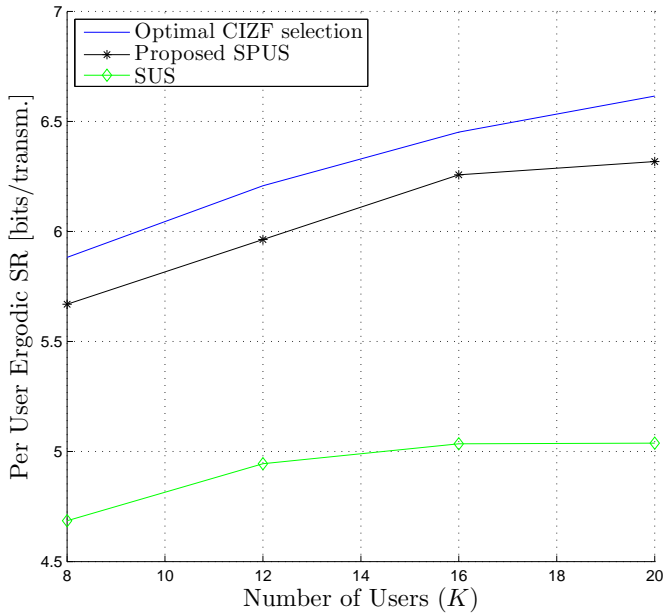}\\
  \caption{Performance of user selection algorithms with respect to the total available transmit power. Results  for $N_t = 4$ users selected out of a variable pool of users, for a fixed total transmit power of $15$ dB. }
\label{fig: GlobeCom 2}
\end{figure}

In Fig. \ref{fig: GlobeCom 2} the performance of the discussed algorithms is  studied with respect the size of the available user set. The beneficial effect of the increasing number of users is clear for all algorithms. What is more, for relatively small user pools the performance of the algorithms is beginning to saturate thus indicating that the main gains are  gleaned for finite numbers of users, that are in line with the dimensions of practical operating multiuser systems.

%

 \section{Conclusions and future work }\label{sec: conclusions}
The concept of constructive interference in linear precoding systems has been examined under the framework of power optimization for the maximization of the system performance. Results indicate that an excess of 2dB gain can be gleaned by optimizing the power allocation with the aim of increasing the total system throughput in constructive interference precoding systems. Moreover, as it has been shown, the individual user power consumption of these schemes can be further reduced, thus leading to some gains. Finally, the user selection problem has been tackled for the novel type of precoding and a heuristic, low complexity, iterative algorithm with close to optimal performance has been proposed.

Future extensions of this work include  the investigation of constructive interference amongst users with different constellations in an adaptive modulation environment where the  limitations induced by these practical constellations are alleviated.    \section*{Acknowledgment}
This work was   partially supported by the National Research Fund, Luxembourg under the project  ``$CO^{2}SAT:$ Cooperative \& Cognitive Architectures for Satellite Networks'.

\begin{algorithm}\label{fig: SPUS}
\SetAlgoLined
\textbf{Semi-Parallel User Selection (SPUS) algorithm}\\
\KwOut{ $\mathbf{G}_{out} $}
\KwIn{ $\mathbf G = \text{diag}(\mathbf s)\cdot \mathfrak{Re}(R)\cdot \text{diag}(\mathbf s), $ }
\emph{Step 1: Initialization}
 $\pi_{(0)} = \arg \max ||\mathbf h_{k}|| = \arg \max[\mathbf{G}]_{kk}$, \\
$\forall k=1,...M:\mathbf c _{(0)} =  \mathbf G(\pi_{(0)}, k) $, $\mathcal{S}_{(0)} = \pi_{(0)}$ \\ 
 $\mathcal{T}_{(0)}=\{1,\dots K \} - \{ \pi_{(0)}\}\ $ set of unprocessed users.\\
 
\For{$i = 1 \to M$ }
{
\emph{Step 2: Selection}
$\pi_{(i)} = \arg \max  \mathbf{c} _{(i-1)} , \text{ Provided that } \pi_{(i)} \in \ \mathcal{T}_{(i-1)}$;   \\
$\forall k=1,...M: \ \mathbf c _{(i)} =  \mathbf G(\pi_{(i-1)}, k)+ \mathbf G(\pi_{(i)}, k);$\\
 $\mathcal{T}_{(i)} = \mathcal{T}_{(i-1)}- \{\pi_{(i)}\};$\\
 $\mathcal{S}_{(i)} = \mathcal{S}_{(i-1)}+ \{\pi_{(i)}\};$
}\emph{Step 3: Output}\\
$\mathbf G_{out} = \mathbf G(\mathcal S_{(M)})$;\\
\For{$m = 1 \to M$}{
\For{$l = 1 \to M$}{
\If {$\mathbf G(m, k)<0$ }{
 $\mathbf G_{out}(m, k)=0$
 }
 }
}
\caption{Semi-Parallel User Selection Algorithm (SPUS)}
\end{algorithm}

\bibliographystyle{IEEEtran}
\bibliography{refs/IEEEabrv,refs/conferences,refs/journals,refs/books,refs/references,refs/csi,refs/thesis}

\end{document}